\newcommand{\nuc}[2]{$^{#1}$#2}
\newcommand{\nucm}[2]{$^{#1}$#2$^m$}

\newcommand{\GV}{$G_\mathrm{V}$}

\newcommand{\Ft}{$\mathcal{F}t$}

\newcommand{\qec}{$Q_\text{EC}$}

\documentclass[prl,twocolumn,showpacs,preprintnumbers,amsmath,amssymb,superscriptaddress]{revtex4}
\usepackage{graphicx}
\usepackage{dcolumn}
\usepackage{bm}

\begin{document}

\preprint{APS/123-QED}

\title{\qec~values of the Superallowed $\beta$ Emitters \nuc{50}{Mn} and \nuc{54}{Co} }

\author{T.~Eronen} \email{tommi.eronen@phys.jyu.fi}
\author{V.-V.~Elomaa}
\author{U.~Hager} \altaffiliation[Present address: ]{TRIUMF, 4004 Wesbrook Mall, Vancouver, British Columbia, V6T 2A3, Canada}
\author{J.~Hakala}
\affiliation{Department of Physics, P.O. Box 35 (YFL), FI-40014 University of Jyv\"askyl\"a, Finland}
\author{J.~C.~Hardy}
\affiliation{Cyclotron Institute, Texas A\&M University, College Station, Texas 77843, USA}
\author{A.~Jokinen}
\author{A.~Kankainen}
\author{I.~D.~Moore}
\author{H.~Penttil\"{a}}
\author{S.~Rahaman}
\affiliation{Department of Physics, P.O. Box 35 (YFL), FI-40014 University of Jyv\"askyl\"a, Finland}
\author{S.~Rinta-Antila}
\affiliation{Department of Physics, Oliver Lodge Laboratory, University of Liverpool, Liverpool L69 7ZE, UK}
\author{J.~Rissanen}
\author{A.~Saastamoinen}
\author{T.~Sonoda}\altaffiliation[Present address: ]{Instituut voor Kern- en Stralingsfysica, Celestijnenlaan 200D, B-3001 Leuven, Belgium}
\author{C.~Weber}
\author{J.~\"{A}yst\"{o}}
\affiliation{Department of Physics, P.O. Box 35 (YFL), FI-40014 University of Jyv\"askyl\"a, Finland}

\date{\today}

\begin{abstract}
Using a new fast cleaning procedure to prepare isomerically pure ion samples, we have measured
the beta-decay \qec~values of the superallowed $\beta$-emitters \nuc{50}{Mn} and \nuc{54}{Co} to be
7634.48(7)~keV and 8244.54(10)~keV, respectively, results which differ significantly from the
previously accepted values.  The corrected \Ft~values derived from our results strongly support
new isospin-symmetry-breaking corrections that lead to a higher value of the up-down quark mixing
element, $V_{ud}$, and improved confirmation of the unitarity of the Cabibbo-Kobayashi-Maskawa
matrix.

\end{abstract}

\pacs{21.10.Dr, 23.40.Bw, 27.40.+z}

\maketitle

Precise measurements of superallowed $0^+$$\rightarrow$\,$0^+$ nuclear $\beta$ transitions
yield several important tests of the electroweak Standard Model \cite{Ha05a,Ha05b}, including
the most demanding one available for the unitarity of the Cabibbo-Kobayashi-Maskawa (CKM)
matrix.  These tests have by now reached the $\pm$0.1\% level, with the dominant uncertainty
coming not from experiment but from the radiative and isospin-symmetry-breaking corrections
that must be applied to the experimental results.  Though small, these theoretical
corrections sensitively impact the CKM unitarity test and the limit that it sets on possible
physics beyond the Standard Model.  The measurements we report here constitute a crucial test
of new calculations of the isospin-symmetry-breaking corrections \cite{To08}.

Any $\beta$-decay transition is characterized by an experimental $ft$ value, which depends on
three measurable quantities: the total transition energy, \qec, the half-life of the parent
state, and the branching ratio for the particular transition of interest.  The \qec~value is
required to determine the statistical rate function, $f$, while the half-life and branching
ratio combine to yield the partial half-life, $t$.  Currently, there are thirteen superallowed 
$0^+$$\rightarrow$\,$0^+$ transitions with $ft$ values that have been measured to a precision of
between 0.03 and 0.3\%.  According to the Standard Model, once the calculated transition-dependent
correction terms have been applied to each $ft$ value, the corrected quantities -- denoted
\Ft~ -- should be identical for all cases since the \Ft~value is proportional to \GV$^{-2}$, where
\GV~is the vector coupling constant.  In fact, in 2005 the most important validation of the
existing isospin-symmetry-breaking corrections \cite{tow02} was their success in converting the
substantial scatter in the uncorrected $ft$ values into a remarkable consistency among the
corrected \Ft~values \cite{Ha05b}.

This agreement was somewhat clouded soon after by precise Penning-trap \qec-value measurements
for the $^{46}$V superallowed decay \cite{sav05,ero06b}, which shifted that transition's \Ft~value
more than two standard deviations above the average of all other well-known transitions
and prompted a close re-examination of its isospin-symmetry-breaking corrections.  What
ultimately resulted was a new calculation of those corrections, not just for $^{46}$V but for
the other superallowed transitions as well \cite{To08}.  For the first time, the calculations
included core orbitals, the effects of which were small but sufficient to decrease the \Ft~values
of $^{46}$V, $^{50}$Mn and $^{54}$Co relative to the average.  Impressively, the $^{46}$V
anomaly disappeared, but at the same time the \Ft~values for both $^{50}$Mn and $^{54}$Co were 
shifted down far enough that they now disagreed with the average.

Is this a sign that the new calculations are flawed or does it mean that the accepted \qec~values
of $^{50}$Mn and $^{54}$Co are incorrect just as the $^{46}$V \qec~value had been found to be
incorrect?  Supporting the latter possibility is the fact that the key measurements for $^{50}$Mn
and $^{54}$Co were reported in the same reference \cite{von77} as was the discredited $^{46}$V
measurement.   We settle the question in this report where we present the first Penning-trap
\qec-value measurements for $^{50}$Mn and $^{54}$Co.

All ions of interest were produced at the IGISOL facility \cite{hui04} with 13--15~MeV protons
initiating ($p$,$n$) and ($p$,$p$) reactions on enriched ($>$90\%) \nuc{50}{Cr} and \nuc{54}{Fe}
targets.  Since $^{50}$Mn ($t_{1/2}$\,=\,283\,ms) and $^{54}$Co ($t_{1/2}$\,=\,193\,ms) have
much longer-lived ($>$1 min.) isomeric states at $\sim$200-keV
excitation, in each case the ground and isomeric states were both produced in the former reaction;
the $\beta$-decay daughter, either \nuc{50}{Cr} or \nuc{54}{Fe}, was produced in the latter.  All
recoil ions were thermalized, extracted, re-accelerated and mass separated in a dipole magnet
having a mass resolving power of $\sim$500. 
Ions with the selected mass number, either $A=50$ or $A=54$, were then transported to the JYFLTRAP setup.

\begin{figure}[t]
\includegraphics[width=0.8\columnwidth]{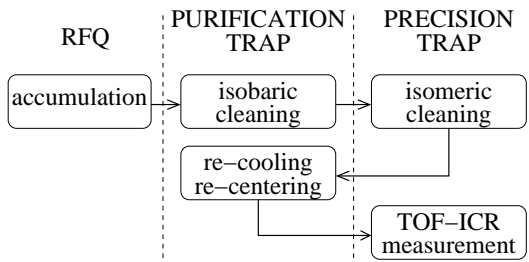}
\caption{\label{fig:ions_go}"Schematic of the full cycle to
perform a cyclotron frequency measurement with
an isomerically clean sample of ions.}
\end{figure}

This setup consists of a radiofrequency quadrupole (RFQ) cooler and buncher \cite{nie02}, which
is used to bunch the beam, followed by two cylindrical Penning traps -- the purification trap and
the precision trap -- housed inside the same superconducting 7-T magnet.  Once a sufficient
number of ions has accumulated in the RFQ, the bunch is transferred to the purification trap
for isobaric cleaning \cite{kol04}.  In our previous \qec-value measurement \cite{ero06b}, 
we successfully used the purification trap alone to prepare isomerically clean samples of
\nucm{26}{Al} and \nuc{42}{Sc} ions.  However, in those cases the difference in cyclotron
frequencies between the ground and isomeric states was $\sim$40~Hz.  For \nuc{50}{Mn} and
\nuc{54}{Co} the separation is only $\sim$10~Hz so we developed a new cleaning scheme
-- visualized in Fig.\,\ref{fig:ions_go} -- in which the ions were transferred to the precision
trap where an electric dipole excitation was applied with time-separated oscillatory fields.
By choosing an appropriate pattern of excitation with time, we could excite ions in the undesired
state to a large orbit while leaving the orbit of the desired ions unaffected.  After this
excitation, the ion sample was transferred back to the purification trap, removing the undesired
ions on the way since they could not pass the 2-mm diaphragm electrode.  The cleaned
bunch was then re-centered in the purification trap and sent again to the precision trap for
its final cyclotron frequency determination. Because this cleaning method is rather fast, 
requiring less than 200~ms to complete, the decay losses were acceptable.

\begin{figure}[b]
\includegraphics[width=\columnwidth]{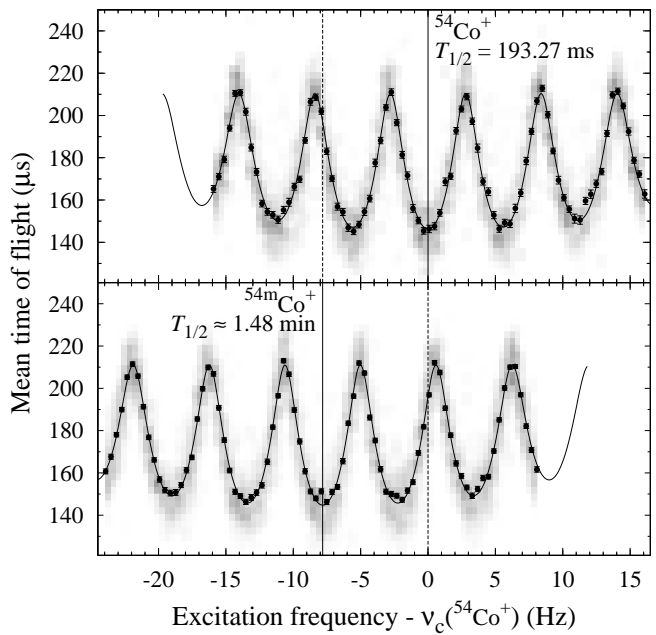}
\caption{\label{fig:ramsey_tofs} Time-of-flight cyclotron resonances obtained using excitation
with time-separated oscillatory fields. An excitation time pattern of 25-150-25~ms (On-Off-On)
was used. The vertical bars denote the cyclotron resonance frequencies, $\nu_c$.  Grey shading
around the datapoints indicates the number of ions in each time-of-flight bin: the darker the grey, 
the more ions.}
\end{figure}

To determine the cyclotron frequency of the ion of interest, we first applied a magnetron
excitation for a short duration in order to establish a magnetron radius of $\sim$0.8~mm. Then
an electric quadrupole excitation was applied to mass-selectively convert the magnetron motion
to cyclotron motion.  Finally, the resonance was detected using the time-of-flight ion
cyclotron-resonance technique \cite{gra80,kon95}.  There is an alternative and more
precise approach and, since
the fitting function for excitation
with time-separated oscillatory fields has recently become available \cite{geo07, kre07}, we
took the opportunity to measure part of our data with this so-called Ramsey excitation scheme.
Examples of time-of-flight resonance curves taken with this scheme for \nucm{54}{Co} and
\nuc{54}{Co} are shown in Fig.\,\ref{fig:ramsey_tofs}. 

\begin{figure}[t]
\includegraphics[width=\columnwidth]{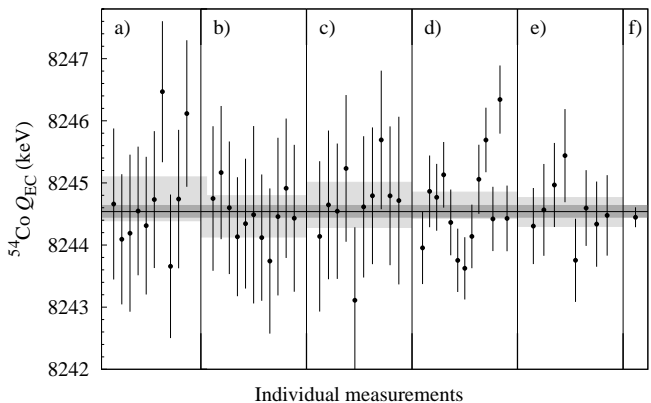}
\caption{\label{fig:54Co_scans}Individual measurements for the \nuc{54}{Co}\,--\nuc{54}{Fe}
\qec~value.  Sets (a) to (c) were obtained with conventional 200-ms rf-excitation but with different
cleaning settings for each set.  Sets (d) and (e) were both obtained with an excitation time
pattern of 25-150-25~ms (On-Off-On) but with different magnetron excitation amplitudes.  The
result labeled (f) is the \qec-value result from the isomer-daughter and isomer-to-ground-state
pairs.  The light-grey bands denote the average of each set; the dark grey band is the final
average including all the data.}
\end{figure}

The \qec~values for \nuc{50}{Mn} and \nuc{54}{Co} were each obtained directly from the frequency
ratio of the mother and daughter nuclei.  As consistency checks, we also measured the
isomer-daughter and isomer-to-ground-state pairs: for example, \nucm{50}{Mn}/\nuc{50}{Cr}
and \nucm{50}{Mn}/\nuc{50}{Mn}.  In all cases, we determined the frequency ratio by interleaving
resonance measurements of one pair member with measurements of the other
until $\sim$10 successive measurements of both had been recorded under identical conditions.  
For every mass pair we obtained several such sets of measurements, each set taken with a
different timing scheme.  We recorded about 4000~ions for each resonance measurement with a
bunch-size distribution maximum kept to 1--2 ions/bunch.  This allowed us to perform a count-rate
class analysis and correct for any possible shift due to contaminating ions \cite{kel03}.  Typical
results are shown in Fig.\,\ref{fig:54Co_scans}.

Since our isomer cleaning technique was completely new, we controlled it carefully and checked to ensure
that it worked properly. For consistency, when measuring a mother-daughter pair we cleaned both the
mother, which required it, and the daughter, which did not.  Then we tested the result by also
measuring the pair with the isomeric state purified not by cleaning but by delaying it several half-lives
for the ground state to decay away.  It was found that if the delay took place in the purification
trap the resonances were of much worse quality, but if it happened in the RFQ trap the quality was
excellent.  The latter result for the resonance frequency agreed well with the result obtained when
the cleaning procedure was applied.

\begin{table}[b]
\begin{center}
\caption{\label{tab:results}Results of the present measurements. ``No." denotes the number of
A-B pairs used in determining the frequency ratio.  The superallowed decay branches are given in boldface. 
The reference mass excesses (Ion B) were taken from Ref. \cite{Au03}.}
\vskip 1mm
\begin{ruledtabular}
\begin{tabular}{lllll}
 & & & \multicolumn{1}{c}{Frequency} & \multicolumn{1}{c}{\qec~or $E_\text{ex}$} \\	
Ion A		& Ion B 	& No. 	& \multicolumn{1}{c}{ratio, $\frac{\nu_\text{B}}{\nu_\text{A}}$}
 &  \multicolumn{1}{c}{(keV)} \\
\\[-3mm]
\hline
\\[-3mm]
{\bf \nuc{50}{Mn}}	& {\bf \nuc{50}{Cr}}	& {\bf 42 }	& ${\bf 	1.0001640971(21)}$	& {\bf	7634.44(10)} \\	
\nucm{50}{Mn}			& \nuc{50}{Cr}			& 58			& 			1.0001689412(13)	& 			7859.81(6) \\
\nucm{50}{Mn}			& \nuc{50}{Mn}			& 38			& 			1.0000048413(20)	&  		225.28(9)	\\
\multicolumn{4}{l}{ Final superallowed \nuc{50}{Mn}\,--\nuc{50}{Cr} \qec} & {\bf 7634.48(7)} \\
\\[-1mm]
{\bf \nuc{54}{Co}}	& {\bf \nuc{54}{Fe}}	& {\bf 52}	& {\bf 1.0001640914(25)}	& {\bf 8244.59(13)} \\	
\nucm{54}{Co}			& \nuc{54}{Fe}			& 55			& 1.0001680222(17)	& 8442.09(9)	\\	
\nucm{54}{Co}			& \nuc{54}{Mn}			& 39			& 1.0000039330(27)	& 197.64(13)	\\
\multicolumn{4}{l}{ Final superallowed \nuc{54}{Co}\,--\nuc{54}{Fe} \qec} & {\bf 8244.54(10)} \\
\end{tabular}
\end{ruledtabular}
\end{center}
\end{table}

Since all our measurements were of mass doublets with the same $A$, any uncertainty
arising from mass-dependent frequency shifts was negligible.  Also, since we interleaved the
measurements of each frequency, we eliminated the effects of any linear drift in the magnetic field.
We accounted for possible non-linear drifts by adding a relative uncertainty of $3.2\times10^{-11} \text{ min}^{-1}$
multiplied by the time in minutes between successive frequency measurements \cite{rah07c}. 

We measured every possible pair (mother-daughter, isomer-daughter and isomer-to-ground state) for both
$A=50$ and 54.  This way we could determine the superallowed \qec~values, not only directly but also
via the isomeric states.  As illustrated in Fig.\,\ref{fig:54Co_scans}, several sets of measurements, each
including $\sim$10 pairs of frequency scans, were obtained for each frequency ratio.  The final result was
derived from the weighted mean with the quoted uncertainty always being the larger of the inner and
outer errors \cite{bir32}.  The results for all six pairs are compiled in Table \ref{tab:results}, 
where the final \qec~values for the two superallowed transitions are weighted averages of the direct
mother-daughter frequency ratio and two-step result linked via the isomer.   We have not added an
additional systematic uncertainty since the systematic shift is expected to be common for all ion
species with the same mass number and makes a relatively negligible contribution to the frequency
ratio uncertainty.

As an additional check, we measured under identical conditions -- including Ramsey cleaning
-- the double $\beta$-decay $Q$ value of $^{76}$Ge, which is known to very high precision
from an off-line Penning-trap measurement with SMILETRAP \cite{suh07}.  The details of our
measurement will be published elsewhere \cite{Ra08} but our result, 2039.04(16) keV, agrees
completely with 2038.997(46) keV, the SMILETRAP result.

\begin{figure}[b]
\includegraphics[width=\columnwidth]{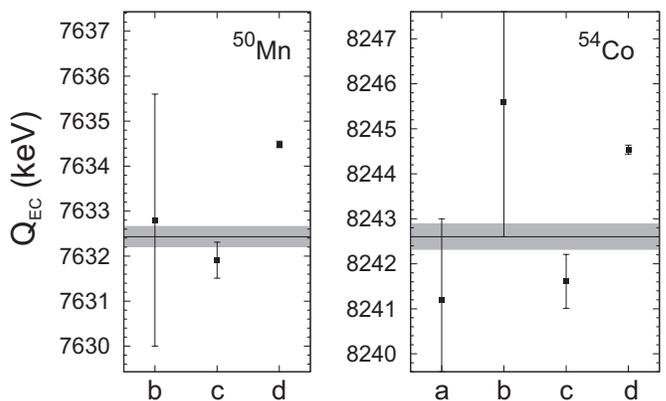}
\caption{\label{fig:compare_to_old}Previous measurements of the \qec~values of \nuc{50}{Mn} and
\nuc{54}{Co} compared to the present results.  The letters on the horizontal scale refer to
the sources: a) Hoath {\it et al.} \cite{hoa74}; b) Hardy {\it et al.} \cite{har74d}; c) Vonach
{\it et al.} \cite{von77}; and d) this work. The grey bar is the value adopted in \cite{Ha05a}, 
including not only the data shown but also the measurements of \qec-value differences \cite{kos87}.}
\end{figure}

In Fig.\,\ref{fig:compare_to_old}, our \qec~values are compared with previous measurements
\cite{har74d,hoa74,von77} and with the average value adopted in the 2005 survey of superallowed
$\beta$ decay \cite{Ha05a}.  Obviously our results are significantly higher than the adopted
averages, principally because the latter were dominated by the measurements published by Vonach
{\it et al.} \cite{von77}, with which we disagree by more than 2.5 keV (5 or more of
their standard deviations).  Evidently, whatever problem Vonach {\it et al.} had with their
measurement of the \nuc{46}{V} \qec~value extended to \nuc{50}{Mn} and \nuc{54}{Co} as well: all
three of these results are lower than the modern more-precise values by approximately the same
amount.

Our results can also be compared with a previous measurement of the difference in \qec~values between
\nuc{50}{Mn} and \nuc{54}{Co}, 610.1(5) keV \cite{kos87}.  Our present results yield the value
610.06(12) keV, in fine agreement.  Less satisfactory is a comparison with the \qec-value difference
between \nuc{42}{Sc} and \nuc{54}{Co}, which was previously determined \cite{kos87} to be 1817.2(2) keV.  
If we use our  present result for \nuc{54}{Co} combined with our recent Penning-trap measurement of
the \nuc{42}{Sc} \qec~value \cite{ero06b}, we obtain a difference of 1818.4(2) keV.  Perhaps the
previous measurement \cite{kos87}, which depended upon ($^3$He,$t$) reactions, included an undetected
target impurity in this case.  

\begin{figure}[b]
\includegraphics[width=\columnwidth]{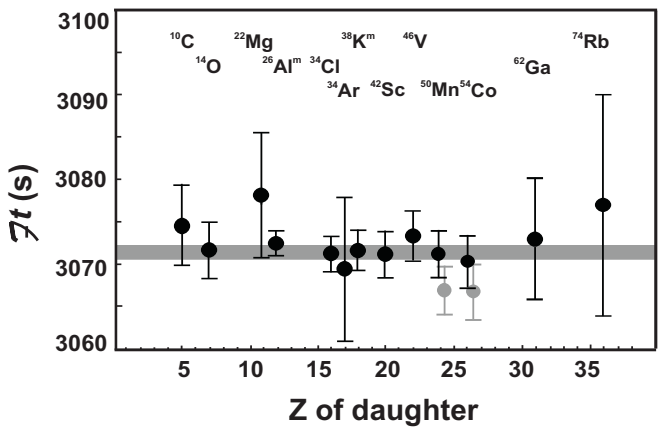}
\caption{\label{fig:Ft_values}\Ft-value results for the thirteen best known superallowed
decays, to which the new isospin-symmetry-breaking corrections \cite{To08} have been applied.  For
the \nuc{50}{Mn} and \nuc{54}{Co} cases, the points shown in grey are the values that were obtained
using the previously accepted \qec~values; those in black result from the new \qec~values reported in
this work.}
\end{figure}

Do our new \qec~values remove the discrepancy between the \nuc{50}{Mn} and \nuc{54}{Co}
\Ft~values and the average \Ft~value for the whole set of thirteen precisely measured superallowed
transitions?  The answer is clearly yes.  The results are shown in Fig.\,\ref{fig:Ft_values}, where
the old values for \nuc{50}{Mn} and \nuc{54}{Co} are shown in grey and our new results are in black.  
In determining the \Ft~values for \nuc{50}{Mn} and \nuc{54}{Co} -- 3071.2(28) and 3070.4(32)\,s, 
respectively -- we combined our new \qec~values
with the half-lives and branching ratios from the 2005 survey of world data \cite{Ha05a}, and applied
the new calculated correction terms reported in Ref.\,\cite{To08}.  The consistency is now excellent, 
an outcome that strongly supports those recent calculations and their inclusion of the effects of core
orbitals.

In supporting the new isospin-symmetry-breaking corrections \cite{To08}, our results also reinforce
the higher value of $V_{ud}$, the up-down quark mixing element of the CKM matrix, that those
corrections led to.  Incorporating our new \Ft~values with the eleven others quoted in
Ref.\,\cite{To08}, we obtain the result that $|V_{ud}|= 0.97408(26)$.  With the values of the other
two top-row elements of the matrix taken from the 2006 Particle Data Group review \cite{PDG06}, the
unitarity sum becomes
\begin{equation}
|V_{ud}|^2+|V_{us}|^2+|V_{ub}|^2=0.9998(10),
\end{equation}
in perfect agreement with Standard Model expectations.

\begin{acknowledgments}
This work was supported by the
EU 6th Framework
programme ``Integrating Infrastructure Initiative - Transnational
Access'', Contract Number: 506065 (EURONS, JRA
TRAPSPEC) and by the 
 Finnish
Center of Excellence Programme 2006--2011 (Project No.~213503, Nuclear and Accelerator
Based Physics Programme at JYFL).
JCH was supported by the U.\,S.~Dept.~of Energy under Grant DE-FG03-93ER40773 and by the
Robert A. Welch Foundation under Grant A-1397.  HP acknowledges financial
support from the Academy of Finland (Project No.~111428). 
\end{acknowledgments}

\bibliographystyle{prsty}
\bibliography{hardy_bib,PRLrefs_long}

\end{document}